## THIN FILMS OPTICAL PROPERTIES, PHOTOSENSITIVE MATERIALS, TRANSIENT ABSORPTION

Correspondence and requests for materials should be addressed to K. V. A. (adarsh@iiserb.ac.in)

**Nanosecond light induced, thermally tunable transient dual absorption bands in a-Ge$_5$As$_{30}$Se$_{65}$ thin film**

Pritam Khan[1], Tarun Saxena[1], H. Jain[2] and K.V. Adarsh[1*]

[1]Department of Physics, Indian Institute of Science Education and Research, Bhopal 462023, India.

[2]Department of Materials Science and Engineering, Lehigh University, Bethlehem, Pennsylvania 18015, USA

**In this article, we report the first observation of nanosecond laser induced transient dual absorption bands, one in the bandgap (TA$_1$) and another in the sub-bandgap (TA$_2$) regions of a-Ge$_5$As$_{30}$Se$_{65}$ thin films. Strikingly, these bands are thermally tunable and exhibit a unique contrasting characteristic: the magnitude of TA$_1$ decreases while that of TA$_2$ increases with increasing temperature. Further, the decay kinetics of these bands is strongly influenced by the temperature, which signifies a strong temperature dependent exciton recombination mechanism. The induced absorption shows quadratic and the decay time constant shows linear dependence on the laser beam fluence.**

Chalcogenide glasses (ChG) are an important class of amorphous semiconductors distinguished from other inorganic amorphous semiconductors by their unique photosensivity depicted in terms of chemical, physical and structural changes upon bandgap/sub-bandgap illumination[1]. It is generally believed that these effects are instigated by the formation of electron-hole pairs in localized band tail states[2-5] having long excitation lifetime and strong electron-lattice coupling[6]. Apart from the fundamental interest, photosensitivity of amorphous ChGs is of great interest to various applications ranging from holographic recording[7], waveguides[8], diffraction elements,[9] nano-antenna,[10] etc.

Light induced shift in the optical absorption of ChG is a phenomenon well studied over many decades. Examples of such effects are photodarkening (PD) and photobleaching (PB) in As and Ge based chalcogenides respectively[11, 12]. PD/PB in ChG is thought to be originating from the strong coupling of photo-generated carriers with lattice through phonons followed by structural changes. Recent, continuous wave (cw) optical pump probe studies on PD/PB revealed that such effects consist of two parts: one is metastable PD/PB and the other is transient PD (TPD) which persists in the medium only during illumination[13-15]. Interestingly, it has been shown that for short illumination with a cw light, transient part is more apparent than the slow and cumulative metastable part[13]. However, most studies on transient effects were conducted with cw light with low time resolution and consequently, an exact assessment of these effects (magnitude, kinetics and spectral width) is lacking. Using transient grating experiments, Rosenblum *et al.*[16] and Regmi *et al.*[17] have demonstrated that ns illumination in $As_{50}Se_{50}$ thin film leads to two kinds of PD effects occurring at fast and slow time scales: the fast process in ns is attributed to carrier diffusion and thermalization, and the slow process in μs is due to the structural changes. However, in all these experiments, the transient effects are probed at a single wavelength and hence information on the induced absorption spectrum by ns excitation is still missing. Likewise, the roles of temperature and fluence of illumination on the induced absorption and kinetics are yet to be established, although they have been apparent for cw illumination.

---

* Author to whom correspondence should be addressed; electronic mail: adarsh@iiserb.ac.in

In this article, we report the 532 nm, 5ns short pulse induced transient absorption (TA) in a-$Ge_5As_{30}Se_{65}$ thin films at various temperatures and fluence of illumination. In contradiction to previous predictions of a broad featureless absorption based on small polaron model[18], we demonstrate that the TA spectrum consists of two thermally tunable distinct absorption bands, $TA_1$ and $TA_2$. The magnitudes of $TA_1$ and $TA_2$ show contrasting characteristics with respect to temperature: $TA_1$ decreases while $TA_2$ increases with rising temperature. Nevertheless, kinetics of such effects becomes faster at high temperatures signifying strong temperature dependent exciton recombination. The magnitude of TA varies quadratically, however, the decay time shows linear dependence on excitation fluence.

**Results**

At first, we have investigated the ns light induced TA (spectral and temporal evolution) of a-$Ge_5As_{30}Se_{65}$ thin film at different temperatures. In this context, changes in absorbance ($\Delta A$) immediately following pump beam excitation (average fluence $\approx$ 37mJ/$cm^2$) at different temperatures are mapped in the contour plot as shown in figure 1 (a –d). It can be seen from the figure that $\Delta A$ (a measure of TA) spectra of the sample consist of two discrete absorption bands $TA_1$ and $TA_2$, at all temperatures of illumination. Existence of these transient dual absorption bands is in contradiction with the previous predictions of a broad featureless absorption based on small polaron model[18]. As can be seen from the figure that, following pump beam excitation, the rise of $\Delta A$ is instantaneous. It decays gradually with time, however the recovery is not complete within our experimental time window of 10 µs. For a better understanding of the temperature dependence, cross-sections of the contour (fig. 1) at 0.5 µs following the pump beam excitation at two contrasting temperatures 15 and 405 K are shown in fig. 2a. Notably, TA bands show a red shift with increase in temperature. This observation demonstrates that the TA bands are thermally tunable. To characterize the temperature dependence of the TA bands further, we have shown in fig. 2b, the ground state transmission spectra of the sample at four different temperatures. It is quite evident from the figure that, similar to TA, ground state transmission spectra also shift to longer wavelength with increasing temperature. The observed redshift in ground state transmission can be explained as a consequence of the change in the degree of deviation from the periodicity of the lattice vibrations with temperature[19]. Quantifying the optical bandgap as the energy of photons where the value of transmittance is 0.1 (absorption coefficient is $10^4$ $cm^{-1}$), the dotted vertical lines in fig. 2a indicate the ground state bandgap of the sample at each temperature. Looking further into figs. 2a and b, we can readily assign the different features of TA. The maximum of $TA_1$ lies in the bandgap region of the sample and therefore, $TA_1$ is ascribed to an interband transition. The origin of such an effect is assumed to be in light induced structural changes in the sample, which is a characteristic of ChG. On the other hand, the maximum of $TA_2$ occurs at much longer wavelengths, corresponding to the sub-bandgap region. Therefore, it is ascribed to localized defect states formed by laser illumination. These defects are assumed to be the valence alternation pair (VAP)[2] and their formation is discussed in the following section. At this point, we envision that the evolution of TA spectra with temperature portrays the distribution of localized state energies within the bandgap and change in their occupancy at the measuring temperatures. To get new insights, we have plotted in fig. 2c the change in the bandgap of the sample as a function of measured temperature. The figure clearly indicates that bandgap decreases with increase in temperature similar to the observation by Cody *et al*.[20] for a-Si-H system. Now, to understand qualitatively how the TA bands may change with temperature, we show in fig. 2d the variation of absorption maxima of $TA_1$ and $TA_2$ as a function of temperature. It can be seen from the figure that $TA_1$ and $TA_2$ exhibit contrasting behaviour with respect to temperature, i.e. $TA_1$ decreases whereas $TA_2$ increases with increasing temperature. Interestingly, at low temperature (15 K) $TA_1$ dominates over $TA_2$ and approach each other as temperature increases. Then at sufficiently high temperature (405 K) $TA_2$ takes over $TA_1$. The overall magnitude of TA (sum of contributions from both $TA_1$ and $TA_2$) is the largest at low temperature.

Next, let us examine the kinetics of TA. Figure 3a shows the temporal evolution of $\Delta A$ at 15 K for the wavelength corresponding to absorption maxima ($TA_1$). As can be seen in the figure, following pump beam excitation, the rise of $\Delta A$ is instantaneous. Its magnitude decays gradually, however the recovery is not complete within the experimental time window of 10 μs. To obtain the decay time constant, we have performed a detailed analysis of the time depndence of the experimental data at selected wavelengths (at which the TA bands show the maximum value). In this context, we have assumed that the excitation of ground state leads to population of a particular state. The change in population of this state can be determined by the following rate equation

$$\frac{dN_i}{dt} = -\sum_{j}^{n} a_{ij} N_i \qquad (1)$$

where $N_i$ and $a_{ij}$ ($i \neq j$) are population density of state i and rate constants for transition from i to j state, respectively. General solution of Eq. 1 is a linear combination of n exponentials and as a result, we have fitted our experimental data using the following equation

$$\Delta A = \sum_{j} A_j \exp{-(\frac{t}{\tau_{ij}})} \qquad (2)$$

where $\Delta A$, $A_i$, t, and $\tau_{ij}$ represent change in absorption of the sample after pump beam excitation, amplitude of the exponential for a particular state, time and decay time constant for transition from i to j state, respectively. Best fit to the experimental data clearly demonstrates that single decay constant is required to quantitatively model the transient absorption data as shown in fig. 3a. For understanding the role of temperature on decay kinetics, we have plotted in fig. 3b, the variation of decay time constants of $TA_1$ ($\tau_{TA1}$) and $TA_2$ ($\tau_{TA2}$) at different temperatures. Clearly, $\tau_{TA1}$ and $\tau_{TA2}$ decrease with increasing temperature, manifesting that the relaxation processes accelerate at higher temperatures i.e. the relaxation of TA is a thermally activated process: the transient defect states associated with it relax immediately at high temperature with the aid of thermal vibration. Note that always $\tau_{TA2} < \tau_{TA1}$ which suggests that the photoexcited carriers have longer life time in the deep traps rather than the shallow traps (deep and shallow are defined with respect to the minimum of the conduction band).

After demonstrating the temperature dependence of TA, we examine the effect of pump beam fluence on the TA. At a fixed temperature (in this case room temperature), we have measured the TA at five different fluences as shown in fig. 4a. The magnitudes of TA show many fold increase with increasing fluence, and the effect is reproducible. In addition, there is a non-reversible component of the TA that also increases, which indicates that the sample undergoes permanent changes (later confirmed with optical absorption measurements). For a better understanding, the values of TA maximum intesnity at different pump beam fluence are shown in fig. 4b. It is apparent from the figure that the observed effect is nonlinear, and a second order polynomial equation fits very well to the experimental data. To estimate the effect of pump beam fluence on the kinetics of TA, we have plotted in fig. 4c the variation of decay time constant at different fluence of the pump beam. It can be seen that the decay time constant scales linearly with a positive slope, i.e. TA is found to decay slower at higher fluence. It indicates that the permanent structural changes become more prominent with increasing fluence, which slows down the relaxation process. This observation was confirmed with the optical absorption measurements, i.e. at higher fluence the sample undergoes permanent PD (fig. 5 a & b). Thus our experimental results reveal that excitation fluence plays a predominant role in controlling the magnitude as well as the kinetics of TA spectra.

**Discussion**

To understand the observed effects, we assume that the variation of magnitude and kinetics of TA with temperature is a result of the temperature dependent non-radiative recombination of electron-hole pairs through transient self trapped excitons[22]. Naturally, such non-radiative recombination mechanism would, in principle leave the material in a different state than the as-prepared initial state − in most cases this would induce structural changes since the excess energy released in the non-radiative recombination is used to overcome the activation barriers to form the defects. We suggest that in our films the exciton (X) recombination can occur through two different paths: (1) directly to ground state (Z) through radiative path, or (2) via a metastable state (Y) through defect creation. However, the latter is more probable in our samples than the former owing to the low photoluminescence efficiency[23]. The metastable state may consist of oppositely charged defect pairs $D^+$ and $D^-$ commonly known as valence alternation pair (VAP), where D and the superscript denote the chalcogen atom and its charge state with respect to perfect structure,  respectively. The two paths can be well understood from the configurational coordinate diagram[24] shown in fig. 6. In ChG, the non-bonding lone pair orbital posseses strong electron-phonon coupling, which strongly favors the formation of low energy defect pairs. The formation energy ($E_d$) of such a defect pair is significantly less than the bandgap energy ($E_g$) as well as the exciton energy ($E_x$). As a consequence, the formed excitons are highly unstable and interact with the lattice to form a $D^+$ - $D^-$ pair, which exactly corresponds to self trapping of excitons. So the amount of energy released in self trapping is simply ($E_x$ – $E_d$). As this trapping energy is a large fraction of bandgap energy, rapid non radiative recombination occurs to the metastable state via self trapped exciton. The extra energy released when excitons get self trapped is used to modify the local structure, which give rise to the permanent change in optical properties. Moreover, owing to the disorder, there will be a distribution of the energy of the $D^+$ - $D^-$ pair, which results in a variation of thermal energy from one site to another site to reverse the reaction. Thus the trapped state is metastable, and requires some thermal energy to return to the ground state. This explains the observation of larger TA at lower temperatures. The self trapped exciton model is further justified by the observation of faster relaxation dynamics at high temperatures with the aid of thermal vibration, as seen in fig. 3b. Thus our experimental results clearly support the temperature dependent exciton recombination mechanisms, which explain well both the magnitude and kinetics of TA at all temperatures. Further, the strong fluence dependence on TA may be due to the multiphoton effect that aids the process of structural rearrangement[16].

In conclusion, we have demonstrated for the first time that nanosecond laser induced transient dual absorption bands near the bandgap and sub bandgap regions of a-$Ge_5As_{30}Se_{65}$ thin film. These observations demosntrate the disagreement with the previous predictions of a broad featureless absorption based on small polaron model. Interestingly, TA spectrum is thermally tunable, which provides a novel way to controlling such effects and hopefully contribute to their potential applications in chalcogenide photo-resists and pulsed holography. Spectral and temporal behaviour of TA can be explained with temperature dependent self trapped exciton model. In this regard, the magnitude of TA varies quadratically, but the decay time scales linearly with the fluence of the inducing beam.

**Methods**

**Sample Preparation**. The bulk sample of $Ge_5As_{30}Se_{65}$ glass was prepared by the melt quenching method, starting with 5N pure Ge, As and Se powders. The cast sample was used as the source material for depositing thin films on a microscopic glass substrate by thermal evaporation in a vacuum of $5 \times 10^{-6}$ Torr.

**Nanosecond Pump-probe spectroscopy.** For pump probe TA measurements, the pump beam was the second harmonics of a Nd:YAG laser (5ns pulses centred at 532 nm with an average fluence of $37mJ/cm^2$ and having a repetition rate of 10 Hz) used in single shot mode and probed with desired wavelengths selected from a Xenon Arc lamp (120 W) using a holographic grating with 1200 grooves/mm. The delay between the pump and the probe beam was created using a digital delay generator. The probe beam overlapped with the pump beam on the sample, and the change in absorbance ( $\Delta A = -log[I_{es}/I_{gs}]$ where $I_{es}$ and $I_{gs}$ are the transmitted intensities of probe pulses after delay time t following excitation by pump beam and in ground state, respectively) of the probe beam is measured. For all measurements, the sample was kept in vacuum inside an optical cryostat (scan range from 10K to 450K) and the experiments were performed after stabilizing to the desired temperature. To study the effect of temperature, we have performed TA at four temperatures (15, 165, 305 and 405 K).

**Optical Absorption Spectroscopy.** A probe beam of very low intensity white light having wavelength range 500-850 nm was used to detect the changes in the transmission spectra of the sample after illuminating with nanosecond laser.

**Acknowledgments**

The authors thank Department of Science and Technology (Project no: SR/S2/LOP-003/2010) and council of Scientific and Industrial Research, India, (grant No. 03(1250)/12/EMR-II) for financial support. They gratefully acknowledge the US National Science Foundation for supporting international collaboration through International Materials Institute for New Functionality in Glass (DMR-0844014).

**Author contributions**

K. V. A conceived the idea. P. K. made the samples. P. K. and T. S. did the experiments. P. K., H. J. and K. V. A. interpreted the results and wrote the manuscript.

**Additional information**

Competing financial interests: The authors declare no competing financial interests.

**Figure 1 | Contour plot of time resolved transient absorption spectra of a-Ge$_5$As$_{30}$Se$_{65}$ thin film at (a) 15 K, (b) 165 K, (c) 305 K and (d) 405 K, when illuminated with 5 ns pulses of wavelength 532 nm and an average fluence of 37 mJ/cm$^2$.**

**Figure 2| (a) Transient absorption spectra of a-Ge$_5$As$_{30}$Se$_{65}$ thin film that exhibit dual absorption bands, shown here for 15 and 405 K. (b) Transmission spectra of as-prepared a-Ge$_5$As$_{30}$Se$_{65}$ thin film at four different temperatures. Figure clearly indicates that absorption edge shifts towards longer wavelength with increasing temperature. (c) Change in bandgap of the sample as a function of temperature. Clearly, the bandgap decreases with increasing temperature. (d) Variation of the magnitudes of TA$_1$ and TA$_2$ as a function of temperature. Solid lines in the figures are guide to the eye.**

**Figure 3|(a) Time dependence of TA$_1$ maxima at 15 K. Blue hollow circles and red solid line represent experimental data and theoretical fit (equ. 2) respectively. Best fit demonstrate that only one decay constant is required to quantitatively model the experimental data. (b) Variation of decay time constant associated with TA$_1$ ($\tau_{TA1}$) and TA$_2$ ($\tau_{TA2}$) as a function of temperature.**

**Figure 4| (a) Temporal evolution of TA maxima as a function of excitation fluence. (b) Change in TA maxima as a function of fluence at different probe delay time, $\tau_d$. The experimental data fit very well to a second order polynomial. (c) Decay time constant of TA at different excitation fluence, which scales linearly with the fluence of pump beam. Symbols and the solid line in (b) and (c) represent experimental data and theoretical fit, respectively.**

**Figure 5| Transmission spectra of a-Ge$_5$As$_{30}$Se$_{65}$ thin film when illuminated with (a) 25mJ/cm$^2$ and (b) 75mJ/cm$^2$. Note that the sample permanently photodarkens at higher fluence, indicating that permanent changes are prominent at higher fluence.**

**Figure 6| Configurational Coordinate diagram depicting two recombination paths of the excitons: (1) to the ground state directly, and (2) to a metastable state via the creation of defect pair, which requires thermal excitation to return to the ground state.**

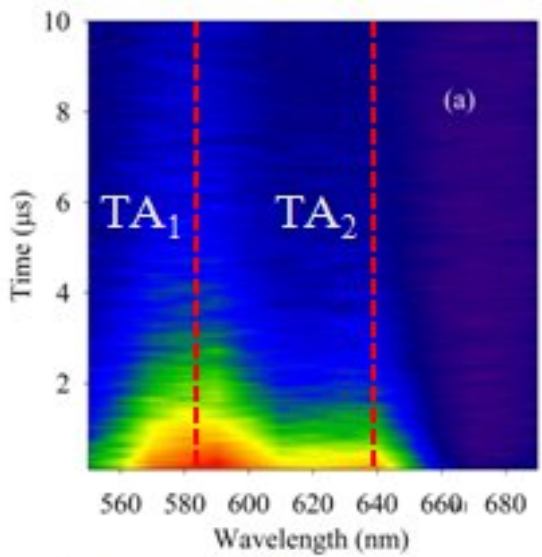
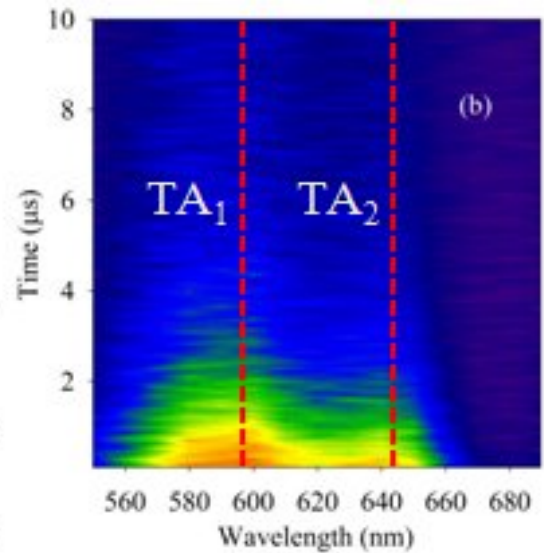
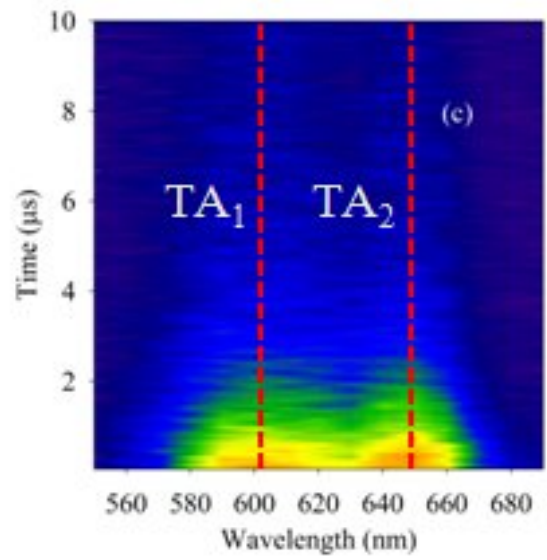
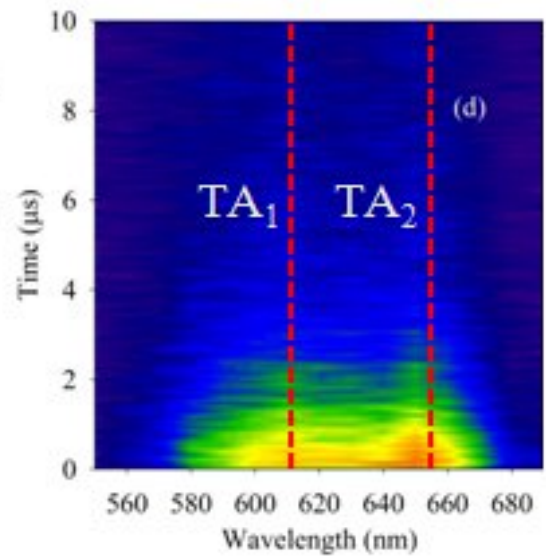

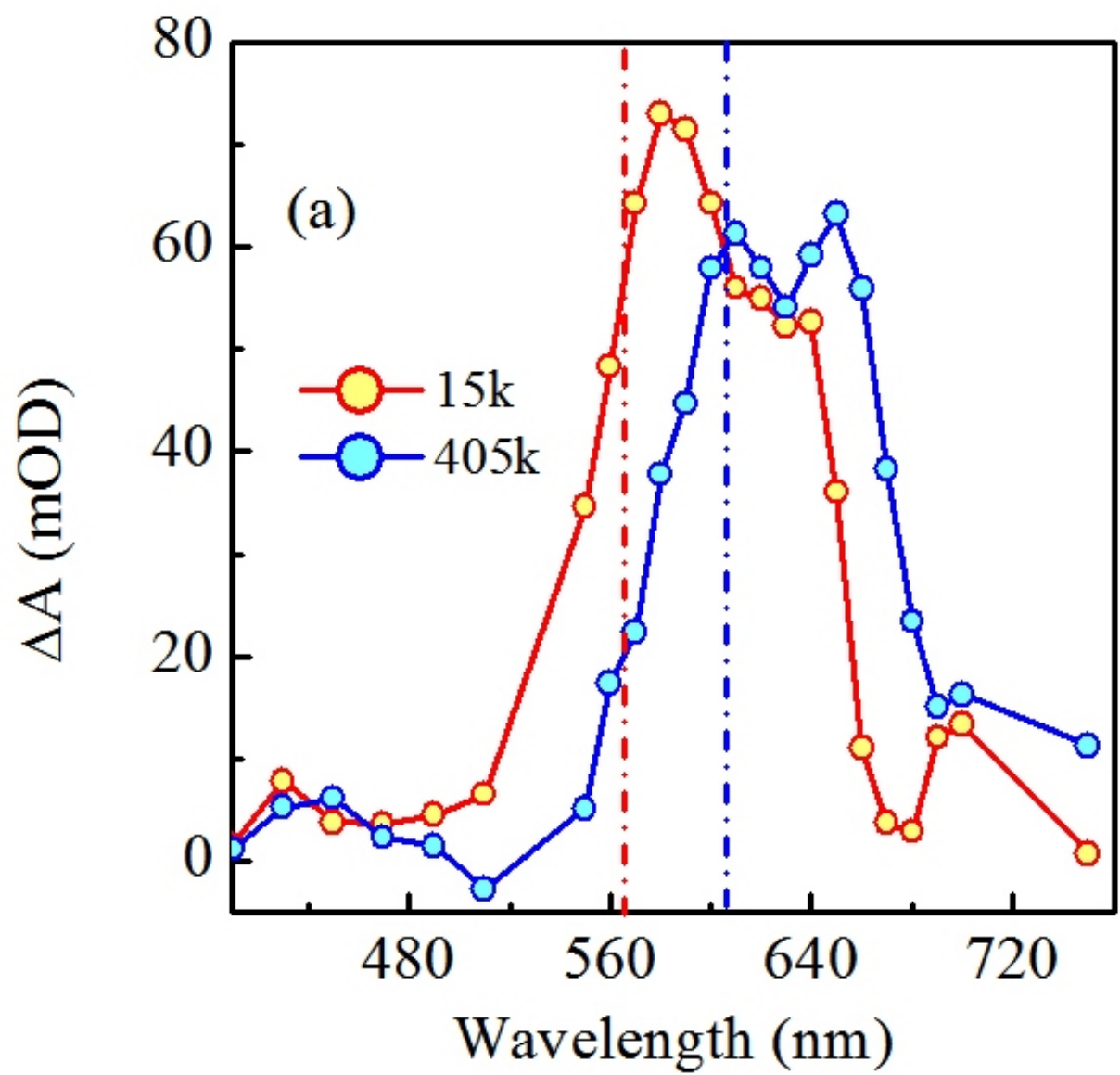
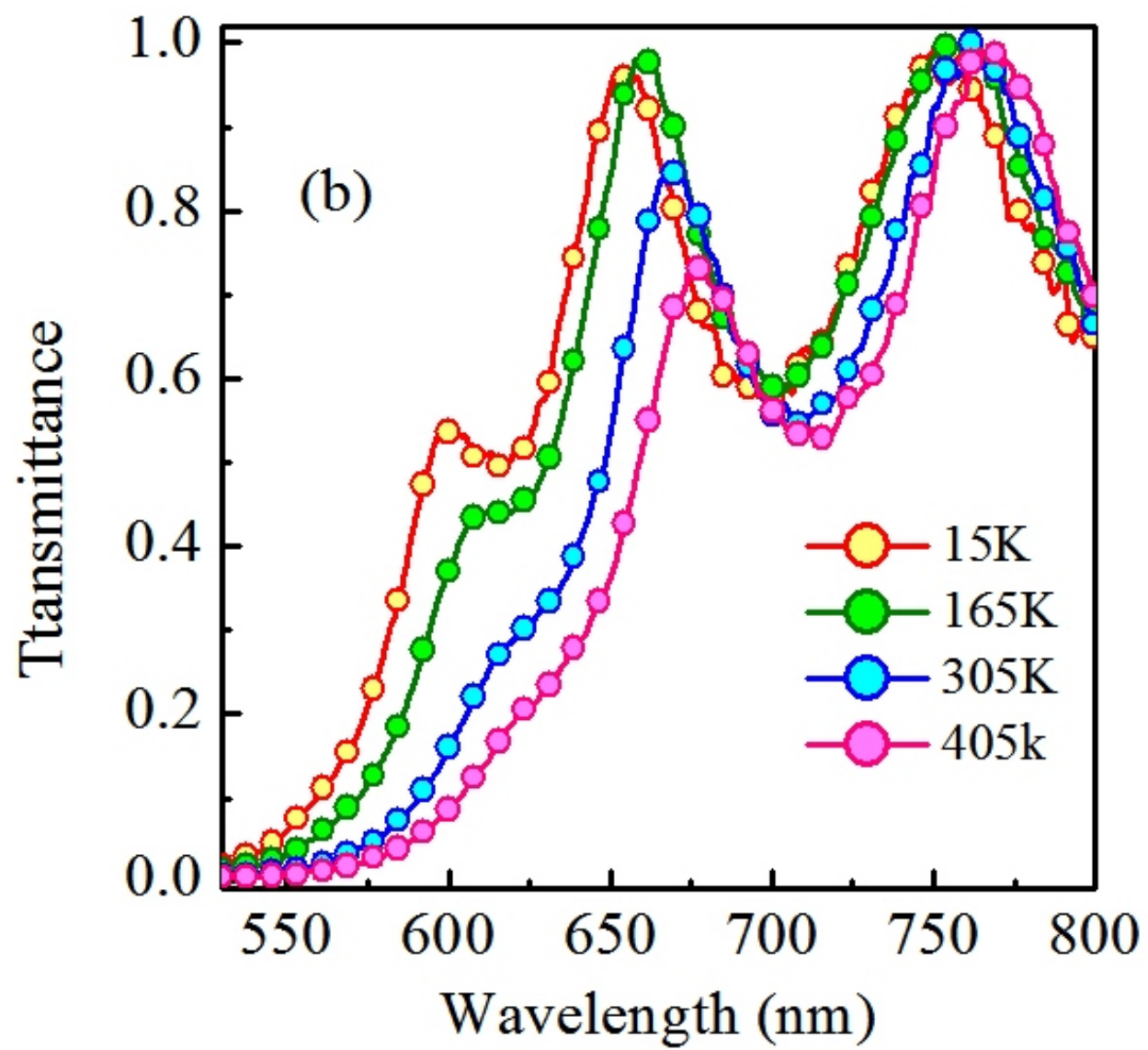
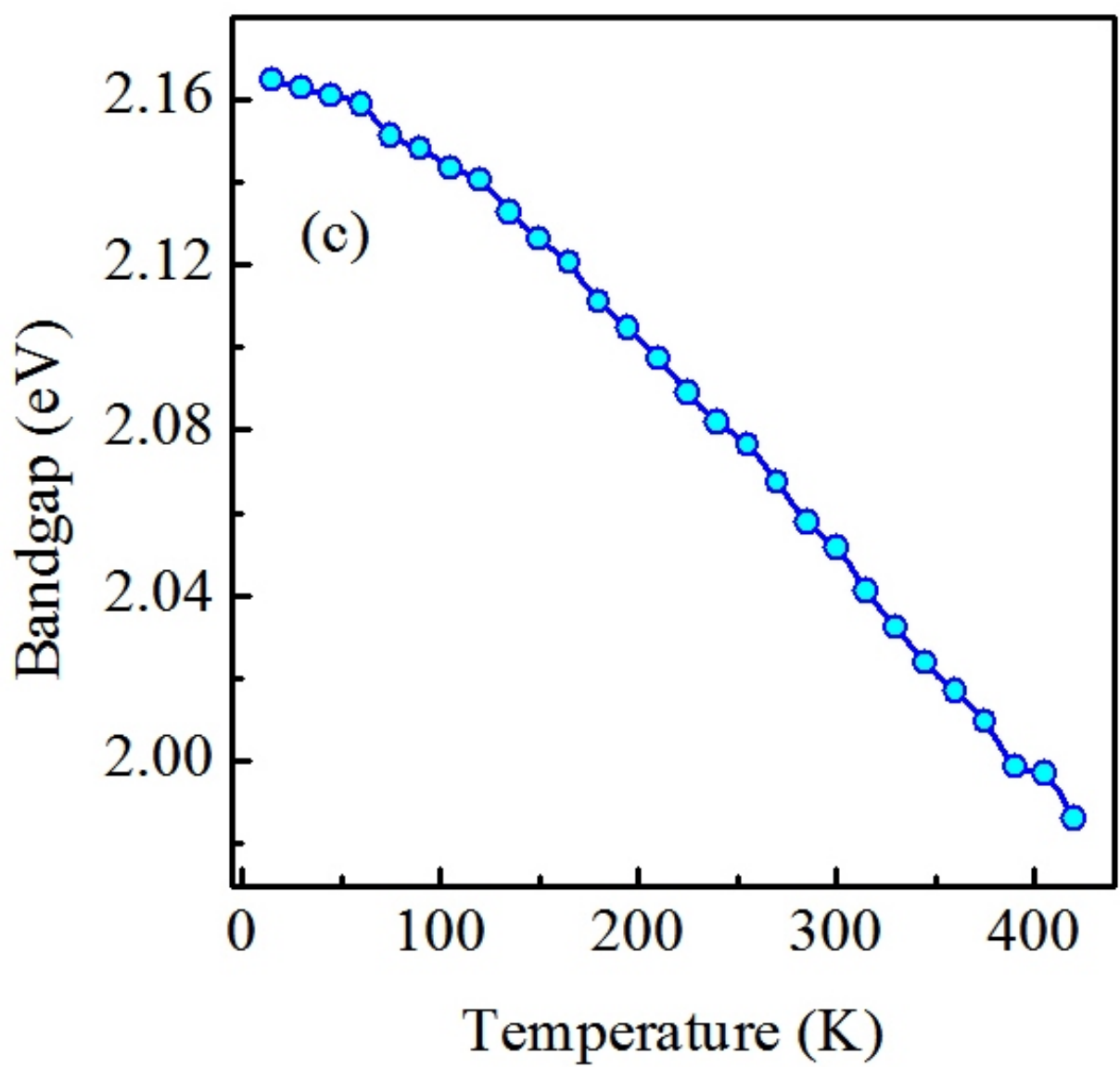
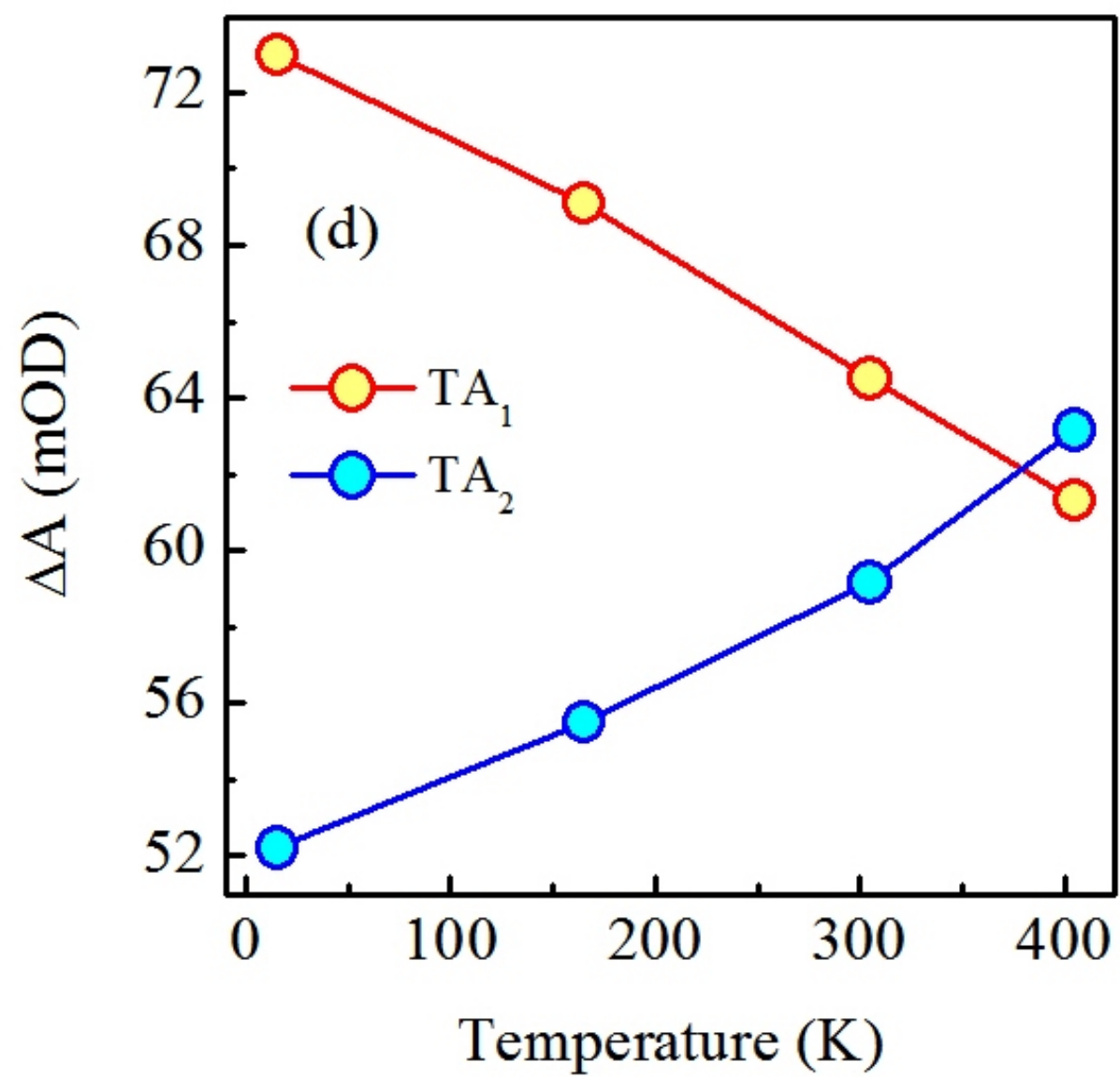

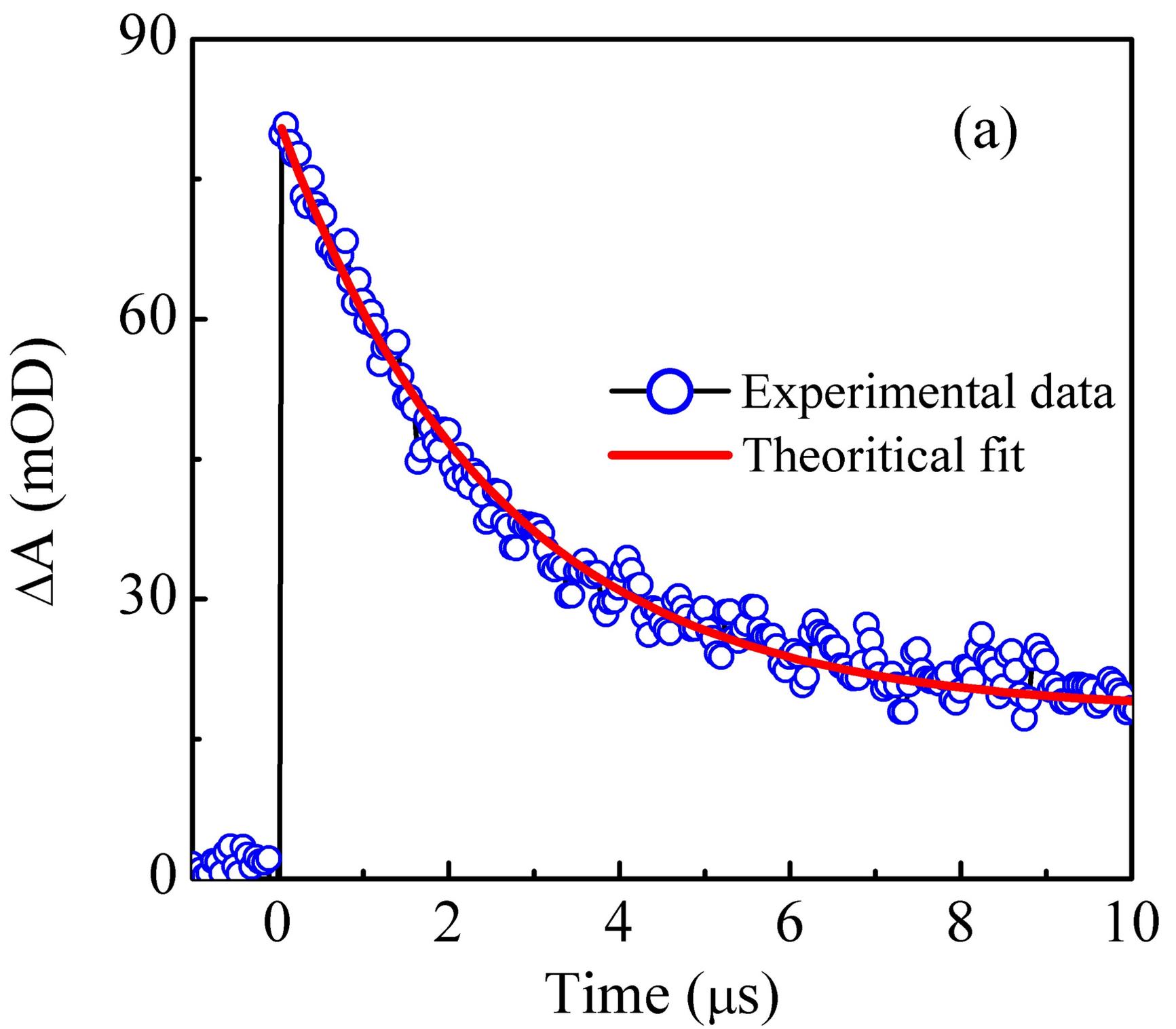

(a)

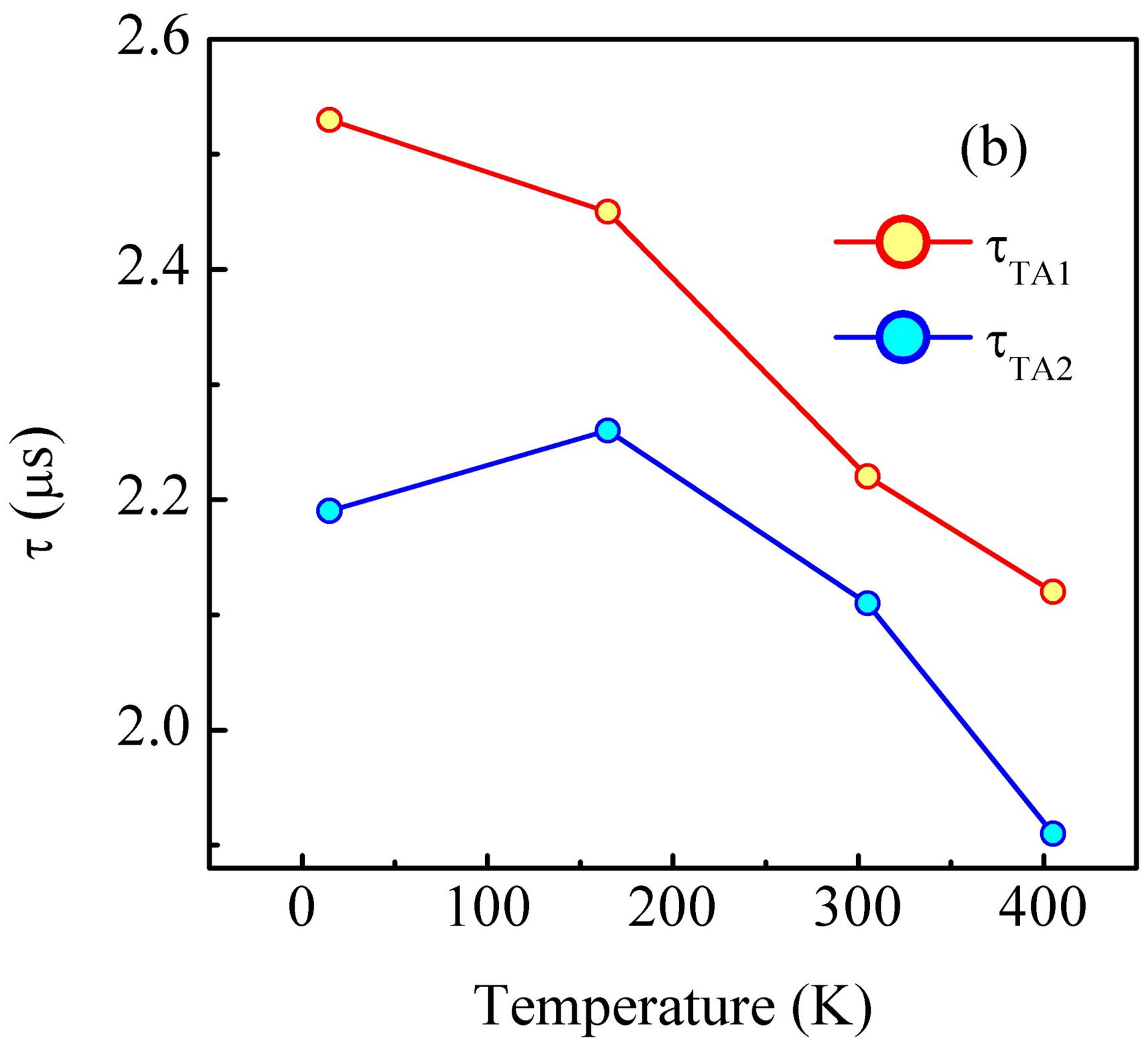

(b)

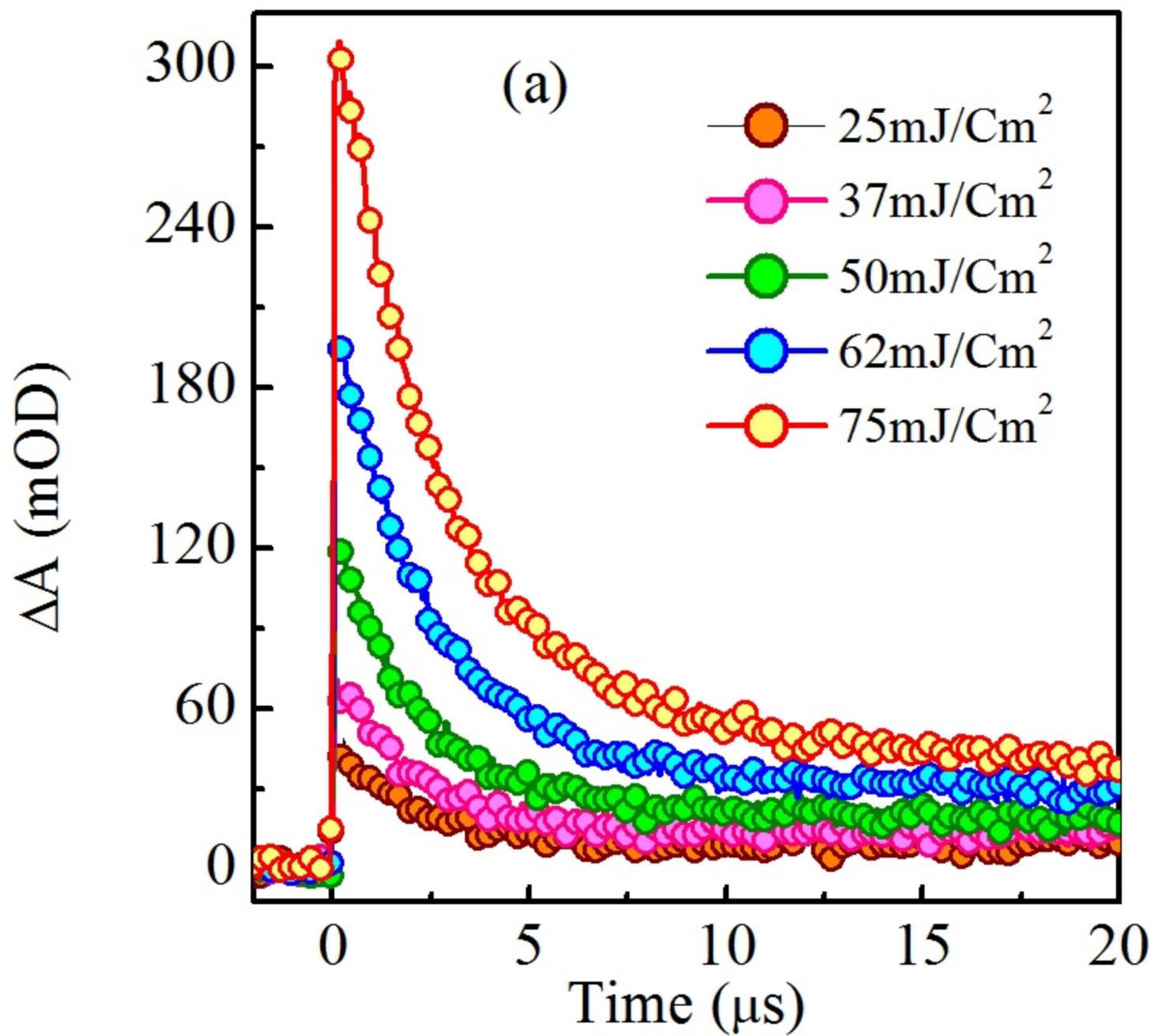

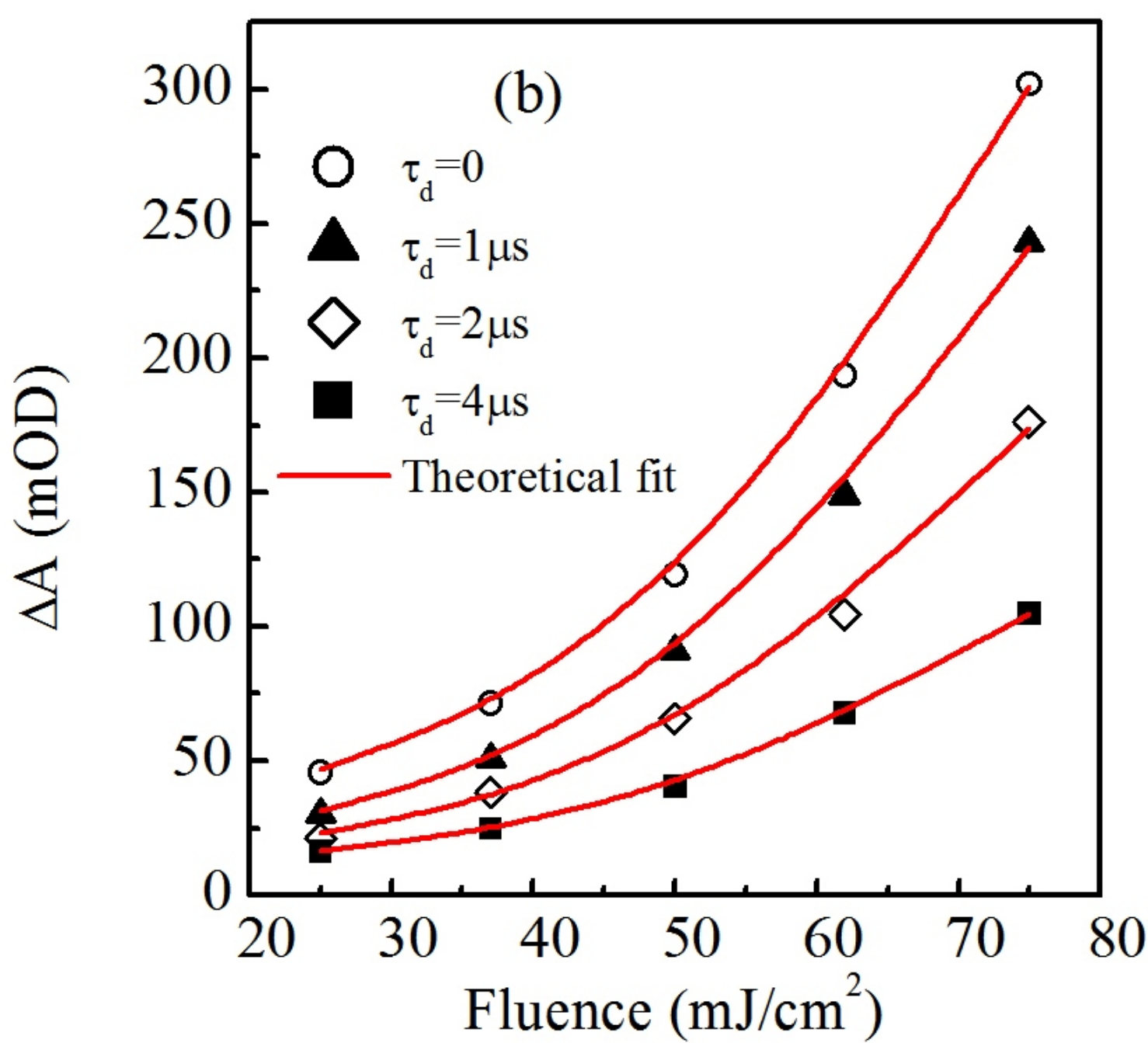

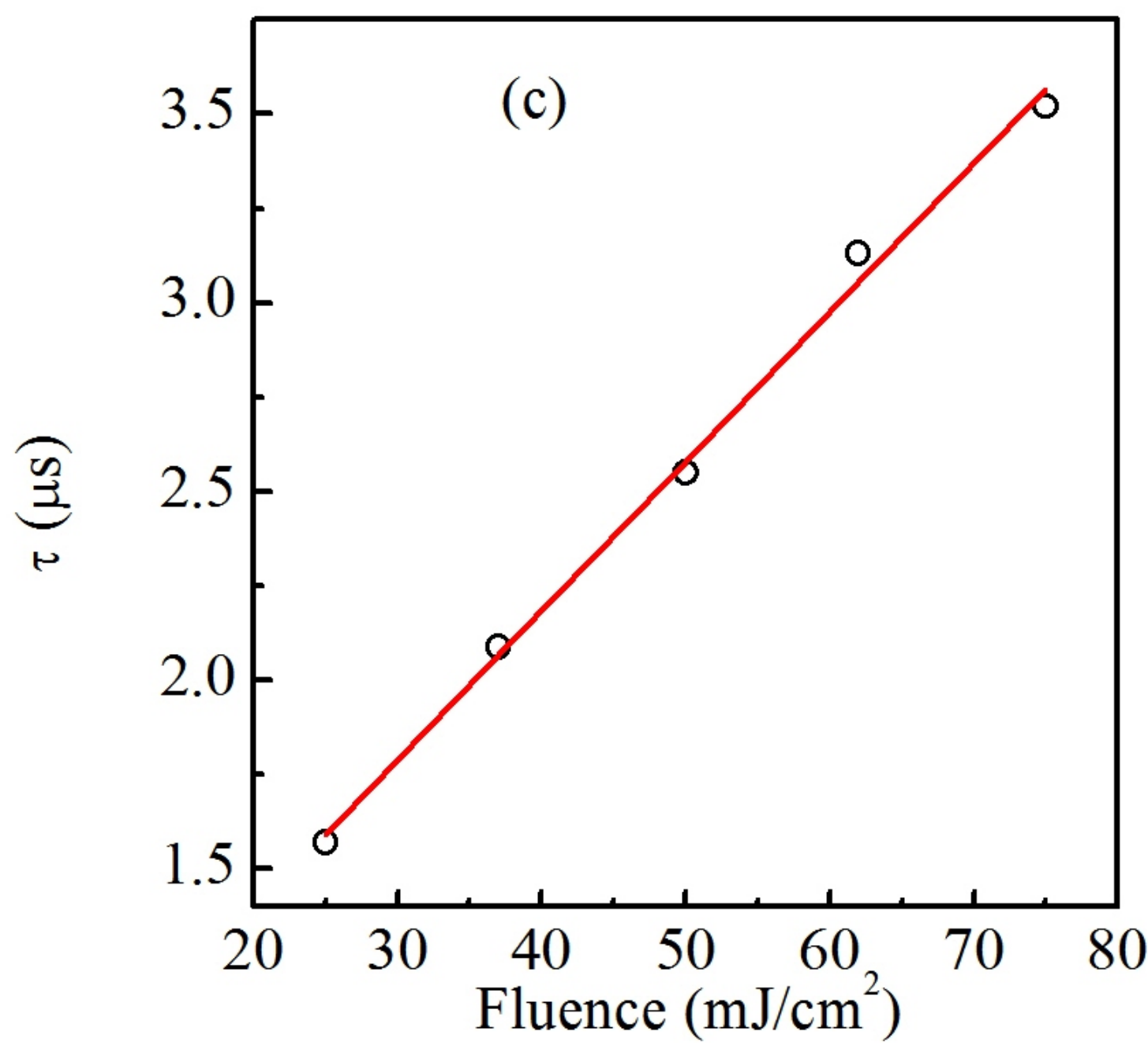

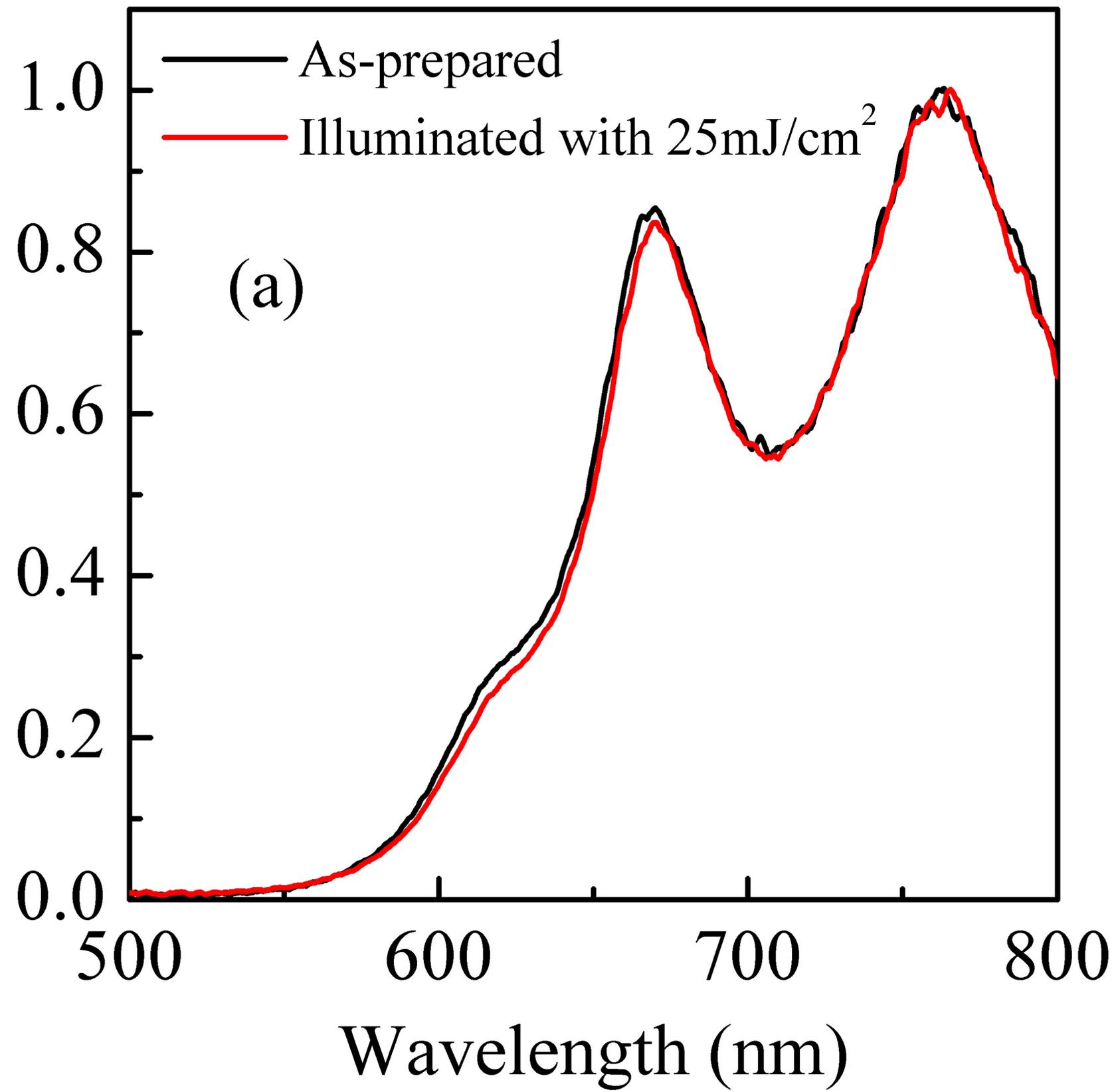

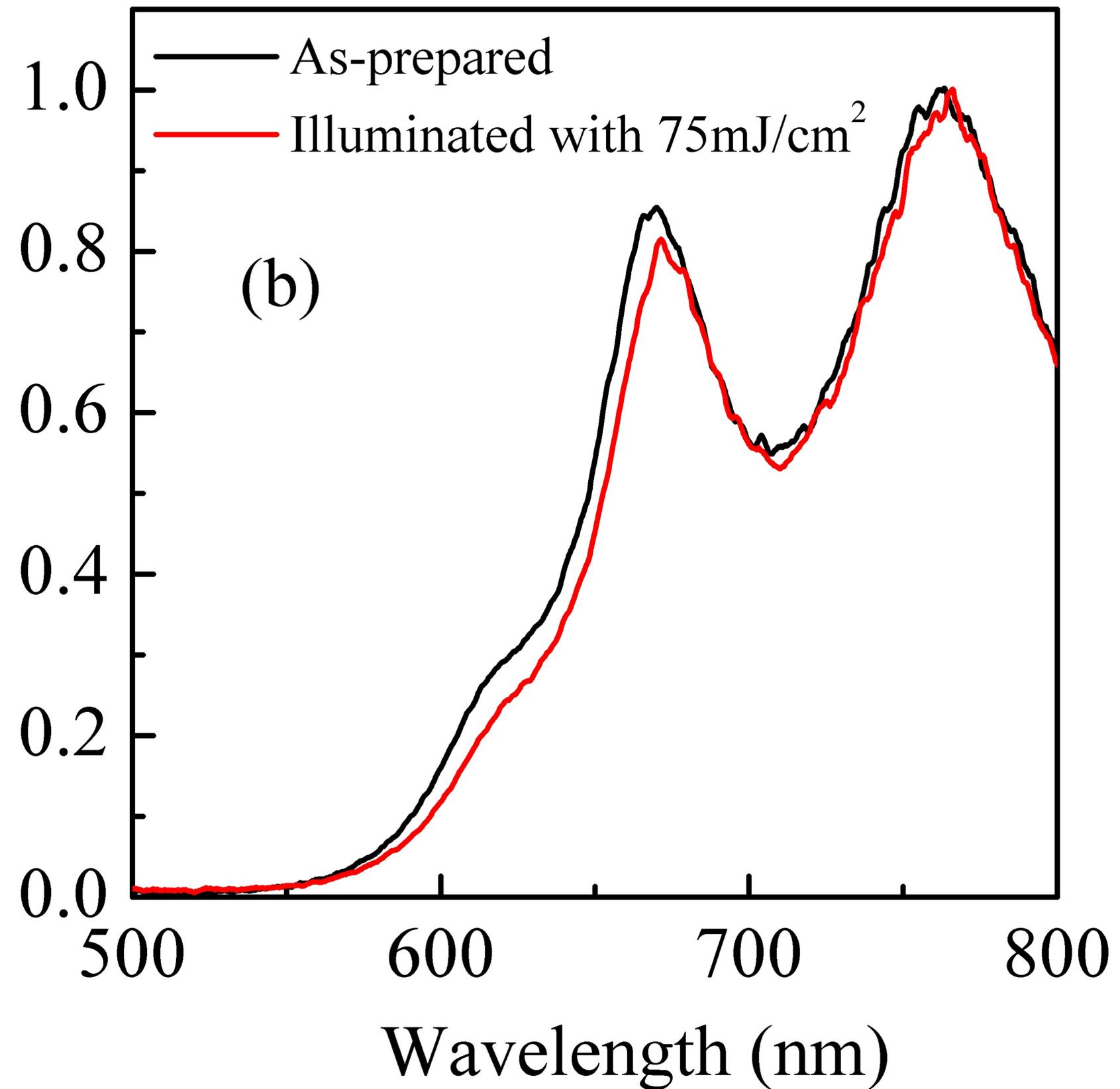

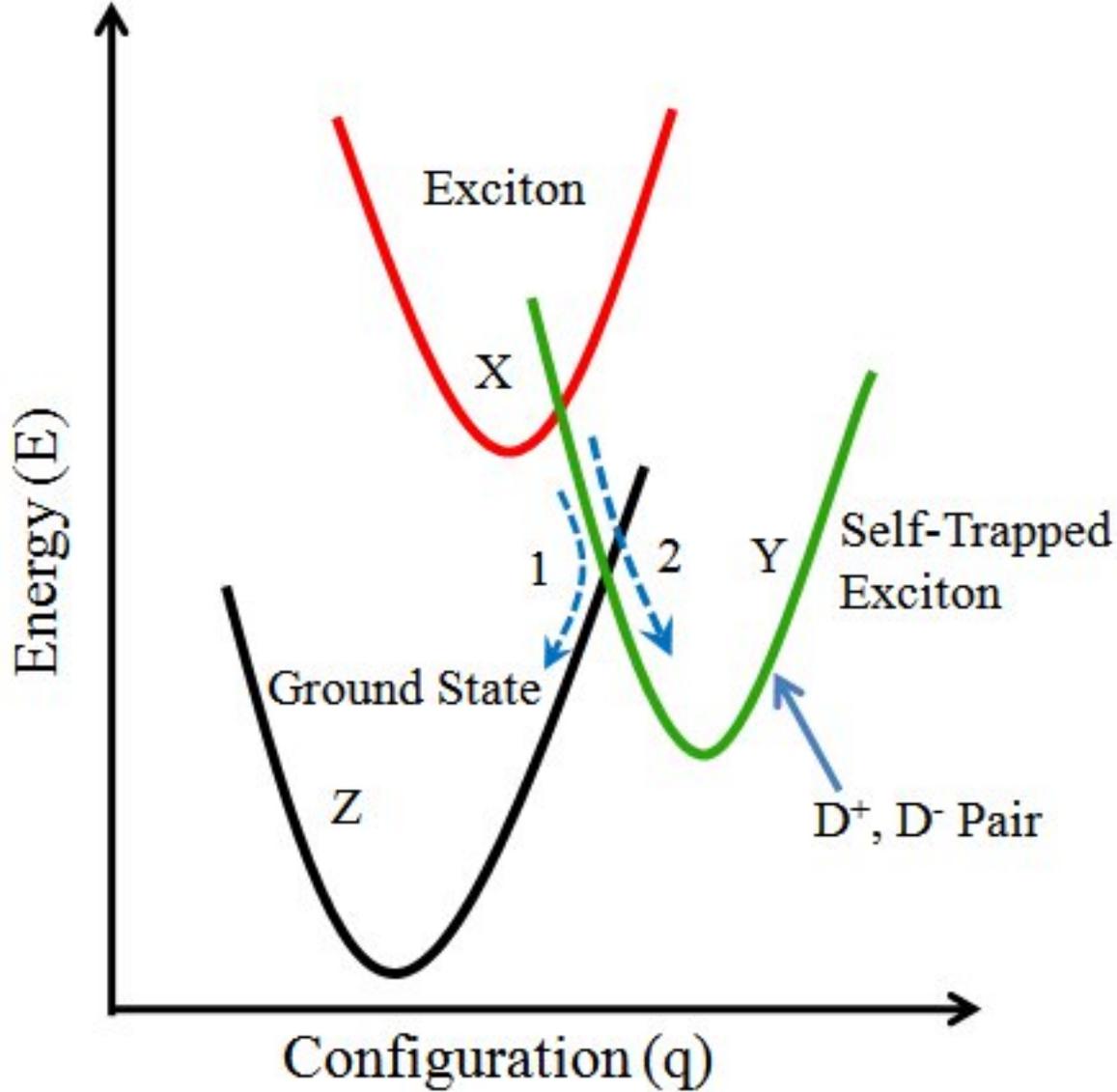